\documentclass[doublecol]{epl2}
\usepackage{amsmath}
\usepackage{graphicx}
\usepackage{color}
\usepackage{hyperref}
\usepackage{psfrag}
\usepackage{stmaryrd}
\usepackage{amssymb}
\usepackage{wasysym}
\usepackage[latin1]{inputenc}

\def \degre {$^\mathrm{o}$}

\title{Earth rotation prevents exact solid body rotation of fluids in the laboratory}
\shorttitle{Earth rotation prevents exact solid body rotation of
fluids}

\author{J. Boisson\inst{1}\thanks{E-mail: \email{boisson@fast.u-psud.fr}}
\and D. C\'ebron\inst{2}\thanks{E-mail: \email{david.cebron@erdw.ethz.ch}}
\and F. Moisy\inst{1}\thanks{E-mail: \email{moisy@fast.u-psud.fr}}
\and P.-P. Cortet\inst{1}\thanks{E-mail: \email{ppcortet@fast.u-psud.fr}}}
\shortauthor{J. Boisson \etal}

\institute{\inst{1} Laboratoire FAST, CNRS, Univ Paris-Sud, UPMC
Univ Paris 06, France\\ \inst{2} Institut f\"ur Geophysik, ETH
Z\"urich, Switzerland }


\abstract{We report direct evidence of a secondary flow excited by
the Earth rotation in a water-filled spherical container spinning
at constant rotation rate. This so-called {\it tilt-over flow}
essentially consists in a rotation around an axis which is
slightly tilted with respect to the rotation axis of the sphere.
In the astrophysical context, it corresponds to the flow in the
liquid cores of planets forced by precession of the planet
rotation axis, and it has been proposed to contribute to the
generation of planetary magnetic fields. We detect this weak
secondary flow using a particle image velocimetry system mounted
in the rotating frame. This secondary flow consists in a weak
rotation, thousand times smaller than the sphere rotation, around
a horizontal axis which is stationary in the laboratory frame. Its
amplitude and orientation are in quantitative agreement with the
theory of the tilt-over flow excited by precession. These results
show that setting a fluid in a perfect solid body rotation in a
laboratory experiment is impossible --- unless tilting the
rotation axis of the experiment parallel to the Earth rotation
axis.}

\begin{document}

\maketitle

\section{Introduction}

There are few examples of fluid mechanics experiments at the
laboratory scale in which the Earth's Coriolis force has a
measurable influence. Such experiments may be considered as fluid
analogues to the Foucault pendulum.  The most popular instance is
certainly the drain of a bathtube vortex~\cite{Perrot1859}.
Although this is the subject of common misconception, it is
actually possible to detect the influence of the Earth's rotation
on the vortex, but only under extremely careful experimental
conditions, far from the everyday experience~\cite{Shapiro1962}.
Thermal convection is another example, in which a slow drift of
the large-scale flow due to the Earth rotation has been detected
in very controlled systems~\cite{Pantaloni1981,Brown2006}.

\begin{figure}
    \centerline{\includegraphics[width=8.5cm]{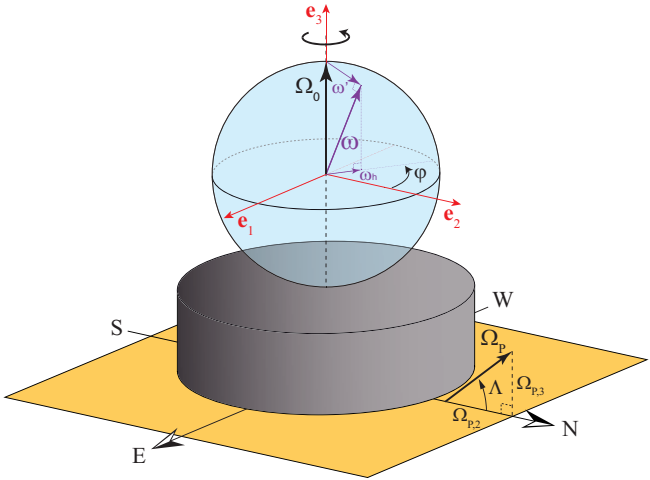}}
    \caption{(Color online) Sketch of the rotating platform and the
    water filled sphere. $({\bf e}_1,{\bf e}_2,{\bf e}_3)$ is a
    Cartesian coordinates system attached to the laboratory frame. The
    platform is rotating at ${\bf \Omega}_0=\Omega_0\,{\bf e}_3$ in
    the laboratory. ${\bf \Omega}_p$ is the Earth rotation vector at
    the latitude $\Lambda=48.70$\degre\, of the laboratory.
    ${\boldsymbol \omega}$ is the rotation vector of the tilt-over
    flow in the bulk. The rotation vectors are not to
    scale.\label{fig:orientation}}
\end{figure}

In this letter we describe an experiment which may be considered
as the most simple fluid Foucault pendulum: it consists in a
volume of water enclosed in a spherical container spinning at
constant rotation rate $\Omega_0$ (fig.~\ref{fig:orientation}).
After a transient known as spin-up, the water is expected to
rotate as a solid body at the same rate
$\Omega_0$~\cite{Greenspan1968}. The timescale for this spin-up is
classically given by the Ekman time $\tau_E = R\,(\nu
\Omega_0)^{-1/2}$, where $R$ is the radius of the sphere and $\nu$
the kinematic viscosity of the fluid. For a typical laboratory
experiment using water, this timescale is usually of order of a
minute, so after a few tens of minutes a perfect solid-body
rotation should be reached, with the fluid exactly at rest in the
frame of the container. If this simple experiment is performed on
Earth, it is expected that the Earth rotation could prevent from
reaching this idealized solid rotation state~\cite{Vanyo2000,
Triana2012}. A weak secondary flow, known as {\it tilt-over}
flow~\cite{Greenspan1968,Busse1968,Malkus1968}, is induced by the
precession of the rotation vector ${\bf \Omega}_0$ of the
container by the Earth rotation vector ${\bf \Omega}_p$. Seen from
the laboratory frame of reference, the fluid particles rotating at
velocity ${\bf u}_0 = {\bf \Omega}_0 \times {\bf r}$ experience a
Coriolis force per unit mass ${\bf f}_c = - 2 {\bf \Omega}_p
\times {\bf u}_0$. This Coriolis force disturbs the fluid
particles periodically at frequency $\Omega_0$, and tends to
deflect their trajectory towards the plane normal to ${\bf
\Omega}_p$.

Precession driven flows in spherical or spheroidal containers and
in spheroidal shells have received considerable interest since
Poincar\'e~\cite{Poincare1910}, because of their importance to
geophysical and astrophysical flows~\cite{Busse1968,Malkus1968}.
In the case of the Earth, rotating with a period $T_0 \sim 1$~day,
the precession of its rotation axis, at a period $T_p \simeq
26~000$~years, could produce large excursions of the rotation axis
of the liquid core~\cite{Greff1999}. Precession driven flows have
also been proposed by Malkus~\cite{Malkus1968} to contribute to
the generation of planetary magnetic fields, which has been later
confirmed by Kerswell~\cite{Kerswell1996} and
Tilgner~\cite{Tilgner2005}. Kida~\cite{Kida2011} recently proposed
a complete solution for the flow in a rapidly rotating sphere
under weak precession, including a detailed analysis of the
conical shear layers detached from the critical latitudes.

First evidence of a tilt-over flow excited by the Earth rotation
in a laboratory experiment has been reported by Vanyo and
Dunn~\cite{Vanyo2000}, using visualizations by dyes and buoyant
tracers, but without quantitative determination of the tilt-over
flow properties. Recently, Triana \etal \cite{Triana2012} obtained
indirect evidence of this effect, from one-dimensional velocity
profiles in a rotating water-filled spherical shell, 3~m in
diameter, containing an inner co-rotating sphere. However, no
quantitative agreement with the theory of Busse~\cite{Busse1968}
could be obtained in their experiment.

Based on the same idea, we provide in this letter, by means of
particle image velocimetry measurements (PIV), the first direct
visualization of the precession flow driven by the Earth rotation
in a sphere rotating in the laboratory. These measurements are a
technical challenge, because of weakness of the velocity signal of
this tilt-over flow (the fluid rotation axis is tilted by less
than 0.2\degre\, with respect to the sphere rotation axis). A
quantitative agreement with the theory of Busse is demonstrated,
both for the magnitude and the orientation of the secondary
circulation.

\section{Physical origin of the tilt-over flow}

Poincar\'e~\cite{Poincare1910} first analyzed the precession flow
in a sphere in the singular case of a perfect fluid. He showed
that the inviscid solution consists in a solid-body rotation
around an axis parallel to $\bf{\Omega}_p$, but of undefined
amplitude. In the presence of weak viscosity, far from the
boundaries, the tilt-over flow may still be described as a
solid-body rotation, with a rotation vector ${\boldsymbol \omega}$
tilted with respect to ${\bf \Omega}_0$, and stationary in the
{\it precessing frame} (the laboratory frame here). We note in the
following ${\boldsymbol \omega'} = {\boldsymbol \omega} - {\bf
\Omega_0}$ the rotation vector of the fluid in the bulk measured
in the rotating frame.

Remarkably, the presence of viscosity, even weak, drastically
changes the rotation vector of the fluid ${\boldsymbol \omega}$
compared to the inviscid solution of Poincar\'e. The orientation
and amplitude of ${\boldsymbol \omega}$ for a viscous fluid are
now non trivial functions of the Poincar\'e number
$\Omega_p/\Omega_0$ and of the Ekman number $E = \nu / (\Omega_0
R^2)$. In the limit $\Omega_p / \Omega_0 \ll \sqrt{E} \ll 1$, the
rotation vector ${\boldsymbol \omega}$ is almost equal to ${\bf
\Omega}_0$, and the small correction ${\boldsymbol \omega'}$ is
almost normal to ${\bf \Omega}_0$. This tilt-over flow has been
described by Busse~\cite{Busse1968} as one among a dense family of
inertial modes, of eigenfrequency given by $\Omega_0$ (see
Ref.~\cite{Greenspan1968} for a general description of inertial
modes in a sphere). When forced by precession, the magnitude
$\omega'$ of this tilt-over flow can be determined by a simple
balance between the Coriolis torque (in the bulk) and the viscous
torque (at the surface of the container). The Coriolis torque is
of order $\Gamma_c \sim \rho R^4 f_c \sim
\rho R^5 \Omega_p \Omega_0 \cos \Lambda$, with $\rho$ the fluid density
and $\cos \Lambda = |{\bf \Omega}_0 \times {\bf \Omega}_p| / \Omega_0
\Omega_p$. The viscous stress is given
by $\sigma \sim \rho \nu \Delta u / \delta$, where $\Delta u
\simeq \omega' R$ is the small velocity jump between the container
wall and the fluid bulk, and $\delta = (\nu / \Omega_0)^{1/2}$ is
the thickness of the Ekman boundary layer. The resulting viscous
torque is of order $\Gamma_{\nu} \sim R^3 \sigma \sim \rho \nu
\omega' R^4 / \delta$. Balancing the two torques gives the simple
relation
\begin{equation}
\omega'  \sim E^{-1/2} \Omega_p \cos \Lambda.
\label{eq:wh0}
\end{equation}
Although very weak, this tilt-over correction may be significantly
larger than the Earth rotation rate in a typical laboratory
experiment where $E \ll 1$.

\section{Experimental Setup}

\begin{figure}
    \centerline{\includegraphics[width=8cm]{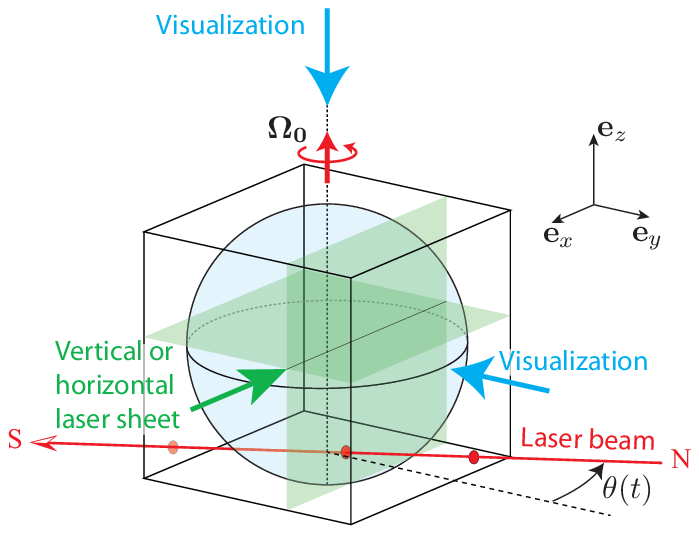}}
    \caption{(Color online) Schematic view of the cubic water tank
    containing the 115~mm radius glass sphere, mounted together on the
    rotating platform. PIV measurements are achieved in off-centered
    vertical and horizontal planes, located at $y_{\rm mes}=+22$~mm
    and $z_{\rm mes}=+22$~mm, using a corotating laser sheet and a
    camera aiming normally at it. The angle $\theta(t)$ between the
    images and the North direction is determined using a continuous
    laser beam aligned along the North-South orientation and crossing
    the rotation axis.\label{fig:setup}}
\end{figure}

The experimental setup, sketched in fig.~\ref{fig:orientation},
consists in a spherical glass tank, of inner radius $R=115\pm
0.25$~mm, filled with water and mounted on the center of a
precision rotating turntable of $2$~m in diameter. We use two
Cartesian coordinate systems, both with origin at the center of
the sphere: (i) $({\bf e}_1,{\bf e}_2,{\bf e}_3)$, attached to the
laboratory reference frame (fig.~\ref{fig:orientation}), with
${\bf e}_1$ pointing to East, ${\bf e}_2$ pointing to North and
${\bf e}_3$ along the vertical; (ii) $({\bf e}_x,{\bf e}_y,{\bf
e}_z)$, attached to the rotating platform (fig.~\ref{fig:setup}),
with ${\bf e}_z={\bf e}_3$, in which the measurements are
performed.

The platform is rotating in the laboratory frame with a rotation
vector ${\bf \Omega}_0=\Omega_0\,{\bf e}_3$. The angular velocity
$\Omega_0$ is varied between 2 and 16~rpm, with temporal
fluctuations less than $\pm 5\times 10^{-4}$. The Ekman number $E
=\nu/(\Omega_0 R^2)$  varies between $3.6\times 10^{-4}$ and
$4.6\times 10^{-5}$ in this range of $\Omega_0$. In
fig.~\ref{fig:orientation}, the rotation vector of the Earth ${\bf
\Omega_p}$ is also shown, for the latitude $\Lambda=48.70$\degre\
of our laboratory in Orsay. The relative scale of the vectors
${\bf \Omega_0}$ and ${\bf \Omega_p}$ is obviously not realistic
in this figure: the Earth rotation rate is $\Omega_p \simeq 6.9
\times 10^{-4}$~rpm $\sim 2\pi/(\mbox{1 day})$, which yields a
Poincar\'e number $\Omega_p / \Omega_0$ ranging from $3.5\times
10^{-4}$ to $4.3 \times 10^{-5}$.

After the start of the platform rotation, we wait at least $\tau_w
= 2$~hours before data acquisition in order to reach a stationary
regime. This waiting time represents at least $30 \, \tau_E$,
where $\tau_E = R \,(\nu \Omega_0)^{-1/2}$ is the Ekman spin-up
time. This indicates that the solid-body rotation state should be
reached, apart from precession effects, with a relative precision
better than $\exp(-\tau_w / \tau_E) \simeq 10^{-13}$.

Velocity fields are measured in the rotating frame  using a
two-dimensional PIV
system~\cite{Davis} mounted on the rotating platform, in either a
vertical $({\bf e}_x,{\bf e}_z)$ or a horizontal $({\bf e}_x,{\bf
e}_y)$ plane (fig.~\ref{fig:setup}). These measurement planes are
off-centered, at $y_{\rm mes}/R = z_{\rm mes}/R \simeq 0.19$ (see
fig.~\ref{fig:setup}), in order to get better insight in the
spatial structure of the flow. Optical distortions are reduced by
immersing the glass sphere in a square glass tank of 300~mm side
also filled with water. The distortion is found less than 5\% for
$r < 0.9R$. The fluid is seeded with $10$~$\mu$m tracer particles,
and illuminated by a corotating laser sheet generated by a
$140$~mJ Nd:YAG pulsed laser. For both horizontal and vertical
measurements, the  sphere cross-section is imaged with a high
resolution $2048 \times 2048$~pixels camera aiming normally at the
laser sheet.

For each rotation rate $\Omega_0$, a set of 2\,000 images is
acquired, covering at least 80 rotation periods. The sampling rate
is synchronized with the platform rotation rate, with a number of
images per rotation ranging from 24 (for low $\Omega_0$) to 9 (for
large $\Omega_0$). PIV fields are computed over successive images
using $32 \times 32$ pixels interrogation windows with $50\%$
overlap, leading to a spatial resolution of about $2$~mm. This
resolution is not enough to resolve the thickness of the Ekman
boundary layers, $\delta \simeq R \, E^{1/2} = 0.8-2.2$~mm, but is
appropriate for the large scales of the precession flow expected
in the bulk.

In view of the very low velocity expected for the precession flow,
the resolution of the velocity measurement is critical in our
experiment. The characteristic velocities of the flow encountered
in this work ranges from $0.01$ to $0.4$~mm~s$^{-1}$ for
$\Omega_0$ between $2$ and $16$~rpm. For the sampling rates
considered here, these velocities correspond to a typical
frame-by-frame particle displacement of 0.16 to 2.6 pixels only.
Although very weak, such displacement may actually be measured
using PIV with sub-pixel interpolation of the correlation peak.
For interrogation windows of size $32 \times 32$ pixels, an
accuracy of 0.05 pixel can be achieved using this
technique~\cite{Raffel2007,Davis}, yielding a signal-to-noise
ratio ranging from 3 (low $\Omega_0$) to 50 (large $\Omega_0$).

The orientation of the experiment with respect to the Earth
rotation axis is monitored using a continuous laser beam aligned
along the North-South direction and passing through the rotation
axis of the sphere (see fig.~\ref{fig:setup}). The beam crosses
the cubic glass tank and is therefore visible on the recorded
images. The angle $\theta(t)$ between the South-North direction
and the measurement fields (see fig.~\ref{fig:setup}) can be
determined for each image with a precision better than $\pm
0.5$\degre.

\section{Structure of the tilt-over flow}

\begin{figure}
    \centerline{\includegraphics[width=8cm]{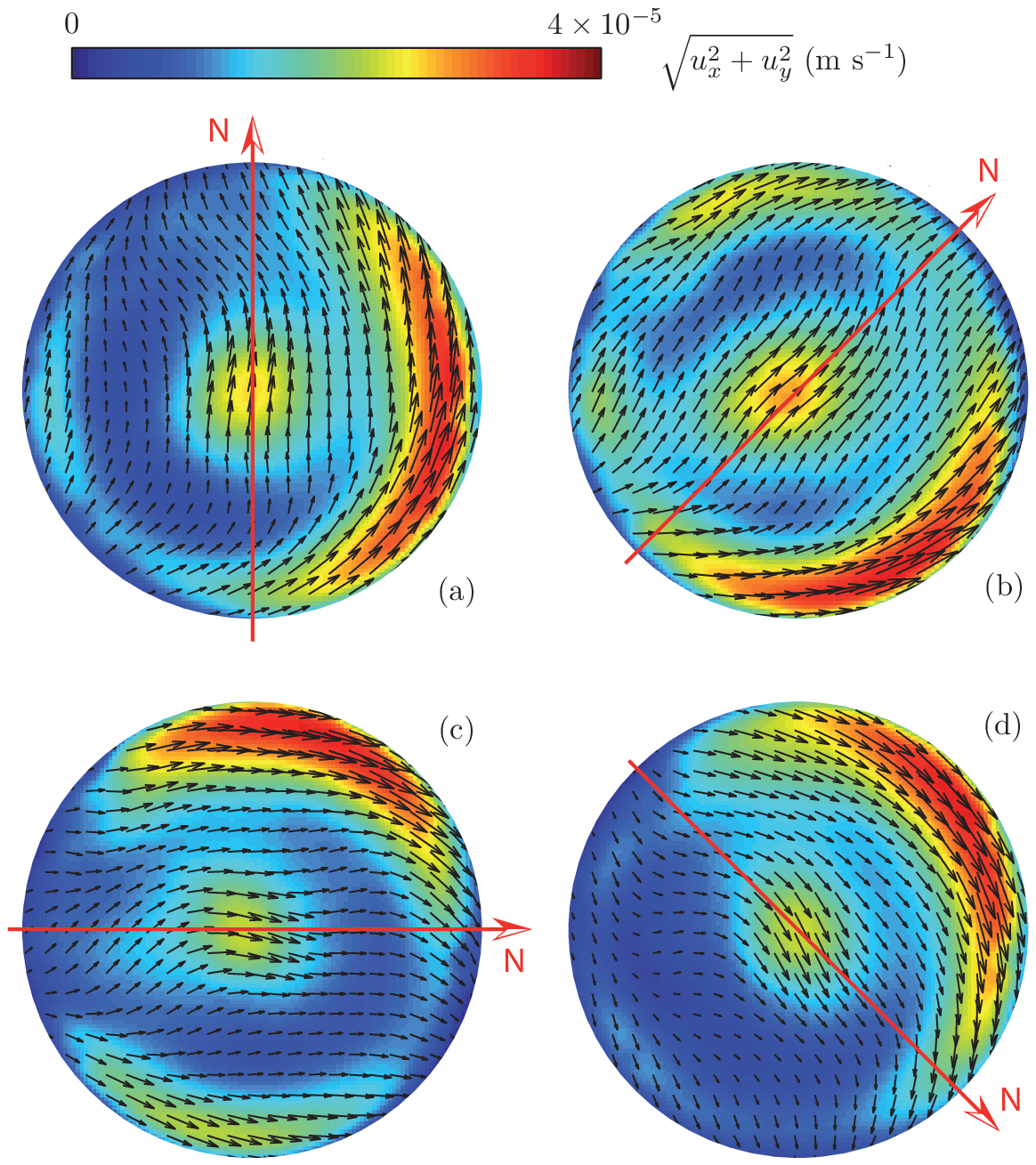}}
    \caption{Horizontal velocity fields measured in the rotating
    frame, in the off-centered horizontal plane at $z_{\rm mes}/R
    \simeq 0.19$ for $\Omega_0=6$~rpm ($E=1.2 \times 10^{-4}$), with a
    phase shift of $\pi/4$ between each image. The platform rotation
    is anticlockwise. The red arrows indicate the direction of the
    North at each time. Resolution of the velocity fields has been
    reduced by a factor 5 for better visibility.
    \label{fig:horizontal}}
\end{figure}

We first show in fig.~\ref{fig:horizontal} the flow measured in
the horizontal plane in the rotating frame for a rotation rate
$\Omega_0=6$~rpm. This flow represents the departure between the
total flow in the laboratory frame and the solid body rotation at
$\Omega_0$. In order to improve the quality of the velocity fields
shown here, a phase average is performed over the velocity fields
at the platform rotation rate $\Omega_0$. This procedure allows to
decrease the broad-band PIV measurement noise by a factor
$N^{1/2}$, where $N$ is the number of recorded rotation periods
($N \geq 80$). The spatial structure of the precession flow can
finally be extracted with a signal-to-noise ratio of at least 30
for all rotations rates.

The four snapshots shown in fig.~\ref{fig:horizontal} are
separated by a phase shift of $\pi/4$,  with a phase origin chosen
such that ${\bf e}_x = {\bf e}_1$ (i.e. ${\bf e}_y$ pointing to
the North). In spite of the very weak velocity signal (of order of
0.04~mm~s$^{-1}$, to be compared to the typical velocity of the
sphere boundaries, $\Omega_0\,R \simeq 72$~mm~s$^{-1}$), we
clearly observe a well-defined flow pattern, which is rotating as
a whole at the platform rotation rate but in the opposite
direction. This weak flow is therefore stationary in the
laboratory frame. Assuming that the total flow in the laboratory
frame is a solid body rotation of vector ${\boldsymbol \omega}$
slightly tilted with respect to ${\bf \Omega}_0$, the measured
flow must be a solid body rotation of rotation vector
${\boldsymbol \omega'} = {\boldsymbol \omega} - {\bf \Omega}_0$.
Since the measurement plane is shifted at $z_{\rm mes}/R \simeq
0.19$, the resulting horizontal velocity field must be uniform in
the bulk, given by ${\boldsymbol \omega'} \times (z_{\rm mes} {\bf
e}_3)$,  and rotating in the anticyclonic direction at frequency
$\Omega_0$, which is precisely what we observe. Snapshots at other
values of $\Omega_0$ show essentially the same flow patterns.

\begin{figure}
    \centerline{\includegraphics[width=8cm]{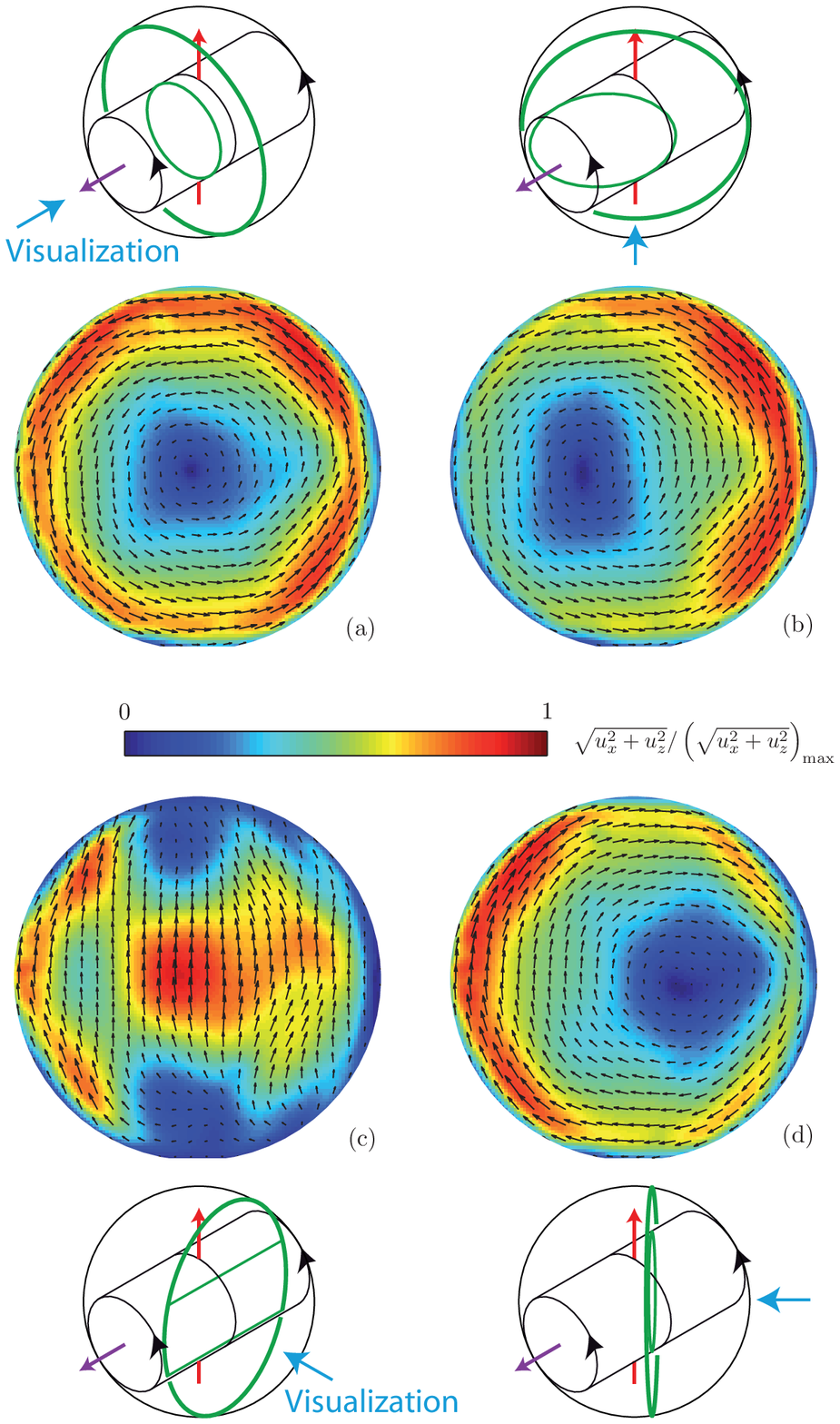}}
    \caption{Vertical velocity fields measured in the rotating frame,
    in the off-centered vertical plane at $y_{\rm mes}/R \simeq 0.19$
    for $\Omega_0=6$~rpm ($E=1.2 \times 10^{-4}$), with a phase shift
    of $\pi/4$ between each image. The color maps the vertical
    velocity norm normalized by its maximum in each field. The phase
    origin is not the same as in
    fig.~\ref{fig:horizontal}.\label{fig:vertical}}
\end{figure}

Measurements in the vertical plane, shown in
fig.~\ref{fig:vertical}, confirm this flow structure. In this
configuration, the camera is now rotating around the vortex of
quasi-horizontal rotation vector ${\boldsymbol \omega'}$
stationary in the laboratory frame. The 4 snapshots taken over
half a rotation around the vortex actually show the following
sequence: (a) anticlockwise, with ${\boldsymbol \omega'}$ pointing
towards the camera; (b) intermediate; (c) ascending, with
${\boldsymbol \omega'}$ pointing to the left; (d) intermediate. If
the tilt-over flow were a pure solid-body rotation, the ascending
flow in the snapshot (c) would be uniform, given by ${\boldsymbol
\omega'} \times ({y_{\rm mes} {\bf e}_y)}$, which is approximately
the case far from the boundaries. The wall region where the flow
departs from a pure uniform flow has a thickness of order of
$0.3\,R$, which is much larger than the expected Ekman thickness
$E^{1/2}R \simeq 0.01\,R$. The tilt-over flow is therefore not
exactly a pure solid body rotation, in agreement with numerical
results obtained in a spherical shell with a very small
stress-free inner solid core~\cite{Busse2001}. Indeed, because of
the breakdown of the Ekman layer at the so-called critical
circles, a pure solid body rotation cannot be a uniformly valid
solution~\cite{Kida2011}.

\section{Viscous prediction for the tilt-over flow forced by precession}

We compute here the rotation vector ${\boldsymbol \omega}$ in the
bulk of the fluid viewed from the precessing frame of reference
(here the laboratory frame), following
Refs.~\cite{Noir2003,Cebron2010}. The differential rotation
between the fluid in the bulk rotating at ${\boldsymbol \omega}$
and the sphere boundary rotating at ${\bf \Omega}_0$ is matched
across an Ekman boundary layer of typical thickness $R E^{1/2}$.
We therefore assume $E \ll 1$, such that a separation between a
bulk flow and a thin boundary layer may be assumed. In the steady
state, the viscous torque ${\bf \Gamma}_\nu$ exerted by the
boundary layers on the fluid in the bulk is balanced by the
Coriolis torque ${\bf \Gamma}_c$ (note that the pressure torque is
zero here because of spherical symmetry). This balance, projected
along $\boldsymbol{\omega}$ and along the two directions normal to
$\boldsymbol{\omega}$, yields the following nonlinear system of
equations~\cite{Greenspan1968,Cebron2010},
\begin{align}
\omega_1^2+\omega_2^2 &= \omega_3(\Omega_0-\omega_3),\label{eq:1}\\
\frac{\Omega_p}{\sqrt{E}}\left(\omega_3 \cos \Lambda -
\omega_2 \sin \Lambda \right) &= \ldots \nonumber\\
& \hspace{-0.5cm} \lambda_r \omega_1 \omega_3^{1/4}
\Omega_0^{3/4} +\lambda_i\omega_2 \frac{\Omega_0^{5/4}}{\omega_3^{1/4}},\label{eq:2}\\
\frac{\Omega_p}{\sqrt{E}} \omega_1 \cos \Lambda &= \lambda_r
\Omega_0^{3/4}\omega_3^{1/4}(\Omega_0-\omega_3),\label{eq:3}
\end{align}
where $\lambda_r$ and $\lambda_i$ are respectively the
non-dimensional viscous damping rate and viscous correction to the
eigenfrequency of the tilt-over mode. Their values have been
obtained by Greenspan~\cite{Greenspan1968} and completed by Zhang
\etal \cite{Zhang2004}, $\lambda_r=-2.62$ and $\lambda_i=0.258$.
In presence of viscosity, the eigenfrequency $\Omega_0$ of the
inviscid tilt-over mode becomes $\Omega_0 + (\lambda_i +
\textrm{i}\lambda_r) \sqrt{E}\sqrt{\Omega_0\,\omega}$
\cite{Noir2003,Zhang2004}, which means that, if the precession
forcing is switched off, the tilt-over mode starts to rotate in
the inertial frame at a frequency $\lambda_i
\sqrt{E}\sqrt{\Omega_0\,\omega}$, while exponentially decaying at
a rate $|\lambda_r|\sqrt{E}\sqrt{\Omega_0\,\omega}$.

Equation~(\ref{eq:1}) reflects the fact the work done per unit
time by the viscous torque is zero, ${\bf \Gamma}_\nu \cdot
{\boldsymbol \omega}=0$, since the work done by the Coriolis force
is zero by definition. This equation, which can be recast into
${\boldsymbol \omega} \cdot ({\boldsymbol \omega} - {\bf
\Omega_0}) = 0$, simply expresses the so-called ``no spin-up''
condition, indicating that there is no differential rotation
between the fluid and the sphere in the direction of the fluid
rotation. This right angle between ${\boldsymbol \omega}$ and
${\boldsymbol \omega'} = {\boldsymbol \omega} - {\bf \Omega}_0$
indicates that the rotation rate $|{\boldsymbol \omega}|$ of the
fluid is lower than $\Omega_0$.

If we further assume that the Poincar\'e number
$\Omega_p/\Omega_0$ is small compared to $E^{1/2}$, the rotation
vector ${\boldsymbol \omega}$ is almost aligned with ${\bf
\Omega_0}$, and the system of equation (\ref{eq:1})-(\ref{eq:3})
can be simplified. More precisely, this regime applies for
rotation rates $\Omega_0 \gg \Omega_{0,c}$, with
\begin{eqnarray}
\Omega_{0,c} = \left(\frac{\Omega_p R \sin \Lambda }{\lambda_r\sqrt{\nu}}\right)^2.
\end{eqnarray}
This condition is comfortably satisfied in the present
experiments, with $\Omega_{0,c} \simeq 5.2 \times 10^{-5}$~rpm. In
this limit, the components of the tilt-over flow can be explicitly
derived,
\begin{eqnarray}
\omega_1 & \simeq & \frac{\Omega_p \cos \Lambda }{\lambda_r} \left( \frac{\Omega_0 R^2}{\nu} \right)^{1/2},\\
\omega_2 & \simeq & \frac{\lambda_i}{\lambda_r} \omega_1,\\
\omega_3 & \simeq & \Omega_0.
\end{eqnarray}
The horizontal projection of ${\boldsymbol \omega}$ in the
laboratory frame, ${\boldsymbol \omega}_h = \omega_1 {\bf e}_1+
\omega_2 {\bf e}_2$, has therefore an amplitude
\begin{equation}
\omega_h  = \frac{\Omega_p \cos \Lambda }{|\lambda_r|} \left(
\frac{\Omega_0 R^2}{\nu} \right)^{1/2} \left( 1+
\frac{\lambda_i^2}{\lambda_r^2} \right)^{1/2},\label{eq:wh}
\end{equation}
which has indeed the expected form~(\ref{eq:wh0}). Note that, in
the limit considered here ($\Omega_p/\Omega_0\ll\sqrt{E}$), the
horizontal projection ${\boldsymbol \omega}_h$ measured in the
experiment almost coincides with ${\boldsymbol \omega'}$.

In this limit, the angle $\varphi$ between ${\boldsymbol \omega}_h$ and ${\bf
e}_1$ (the East direction) is constant, and given by
\begin{equation}
\varphi = \arctan \left( \frac{\omega_2}{\omega_1} \right) = \arctan
\frac{\lambda_i}{\lambda_r} = 174.35^\mathrm{o},
\label{eq:varphi}
\end{equation}
showing that ${\boldsymbol \omega}_h$ points almost to the West
(along $-{\bf e}_1$), with a slight component to the North.
Remarkably, this asymptotic angle obtained in the limit of large
$\Omega_0$ is almost perpendicular to the inviscid prediction of
Poincar\'e, for which  ${\boldsymbol \omega}_h$ points to the
North (i.e. $\varphi = 90$\degre). This indicates that, even for
very low viscosity, the boundary layers have a critical influence
on the tilt-over flow, provided that $\Omega_p / \Omega_0 \ll
E^{1/2}$.

\section{Comparison with the experimental tilt-over flow}

\begin{figure}
    \centerline{\includegraphics[width=8cm]{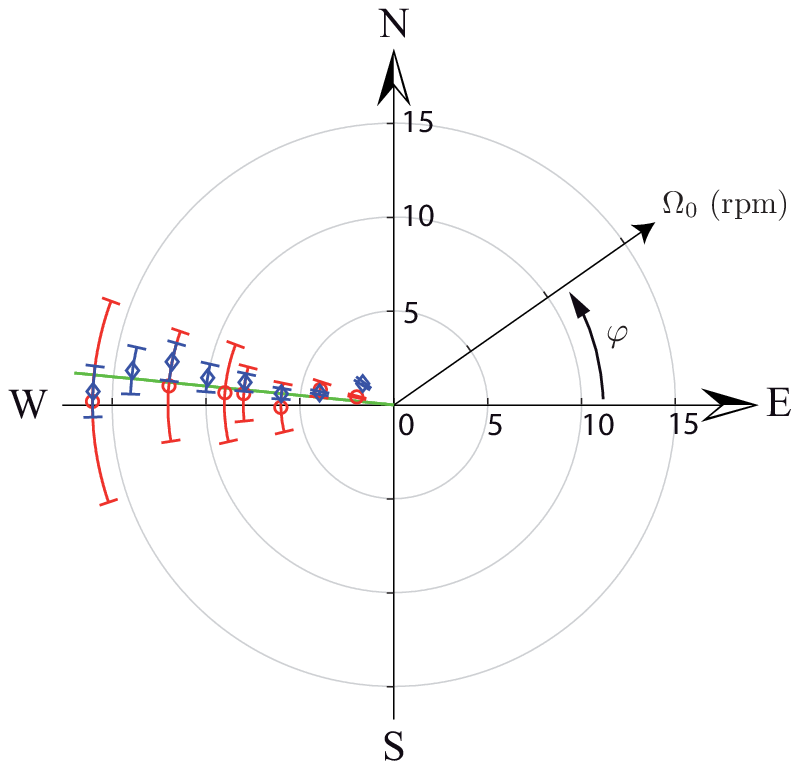}}
    \caption{(Color online) Angle $\varphi$ between the rotation
    vector of the tilt-over flow and ${\bf e}_1$ (i.e. East) as a
    function of the rotation rate $\Omega_0$ in polar coordinates.
    Measurements are obtained in the horizontal ($\circ$) and the
    vertical ($\diamond$) plane respectively. The continuous line
    shows the theoretical prediction $\varphi = 174.35^\mathrm{o}$
    (\ref{eq:varphi}).\label{fig:data}}
\end{figure}

\begin{figure}
    \centerline{\includegraphics[width=7.5cm]{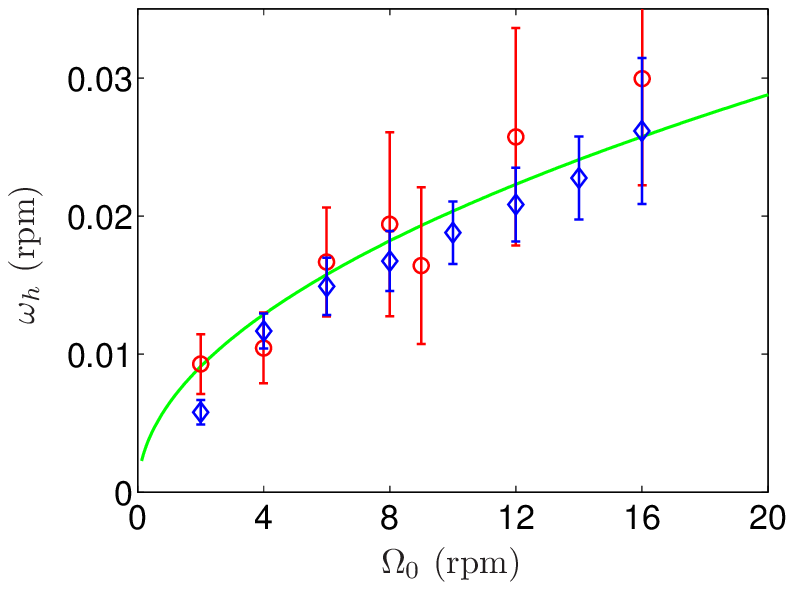}}
    \caption{Horizontal rotation rate $\omega_h$ of the tilt-over flow
    as a function of the rotation rate of the platform $\Omega_0$,
    measured in the horizontal ($\circ$) and vertical ($\diamond$)
    planes. The continuous line shows the prediction (\ref{eq:wh}).
    \label{fig:data2}}
\end{figure}

The rotation rate $\omega_h$ of the tilt-over flow and its angle
$\varphi$ with the East have been systematically determined for
$\Omega_0$ ranging from 2 to 16~rpm. These data have been
extracted independently from the raw velocity fields measured in
the vertical and  horizontal planes, and are compared here with
the theoretical predictions (\ref{eq:wh})-(\ref{eq:varphi}) in
figs.~\ref{fig:data} and \ref{fig:data2}.

Measurements of the vortex angle $\varphi$ from the PIV data in
the vertical plane have been obtained as follows: the horizontal
vorticity, spatially averaged over a central region of 50~mm
radius, shows a harmonic oscillation at frequency $\Omega_0$. At
each period, the delay between the time $t_{\rm max}$ of maximum
vorticity (when ${\boldsymbol \omega}_h$ points to the camera) and
the time at which the North-South laser beam is aligned with the
camera axis is computed. Knowing the instantaneous angle
$\theta(t)$ between the camera incidence and the South-North
direction, we can simply deduce the angle of the vortex as
$\varphi=\theta(t_{\rm max})+90$\degre. An independent estimate
for $\varphi$ has been determined from the data in the horizontal
plane, by computing the time averaged (and spatially averaged over
the region $|{\bf r}|<50$~mm) angle of the velocity with respect
to the East direction ${\bf e}_1$.

The rotation rate $\omega_h$ of the horizontal component of the
tilt-over flow has been determined from the vertical cuts as half
the spatially averaged (over a central disk of radius 50~mm)
vorticity, measured at the times $t_{\rm max}$ of maximum
vorticity. $\omega_h$ has also been determined independently from
the horizontal cuts, as $\omega_h=\langle | {\bf u}_h |
\rangle/z_{\rm mes}$, where $\langle \cdot \rangle$ is an average
over time and over the region $|{\bf r}|<50$~mm, and $z_{\rm mes}$
is the height of the measurement plane.

For both measurements in the horizontal and vertical planes, one
value of $\varphi$ and $\omega_h$ is obtained at each rotation
period. From this set, the average and standard deviation are
computed over the 80 periods recorded for each rotation rate. In
addition to the temporal fluctuations, the errorbars in
figs.~\ref{fig:data}~and~\ref{fig:data2} also include the
variations of $\varphi$ and $\omega_h$ when varying the radius of
the averaging region between $25$ and $75$~mm. For both
quantities, the estimates determined from the two measurement
planes closely agree, although data from the horizontal plane
systematically show a larger scatter.

The vortex angle measured from both vertical and horizontal
planes,  $\varphi \simeq 173 \pm 4$\degre\, and $175 \pm
11$\degre\ respectively (fig.~\ref{fig:data}), are in good
agreement with the theoretical prediction
(\ref{eq:varphi})~\footnote{A possible residual ellipticity of the
sphere would lead to slightly different angles $\varphi$.
Considering a prolate or an oblate spheroid, of ellipticity given
by the maximum deviation of the radius of the sphere ($R = 115 \pm
0.25$~mm), yields predictions for $\varphi$ between 170 and
180$^\mathrm{o}$ for the range of $\Omega_0$ considered here,
which is compatible with the present data.}. Similarly, the
rotation rate $\omega_h$ measured in both planes closely follow
the prediction (\ref{eq:wh}) to within 20\% over the range
$\Omega_0 = 2 - 16$~rpm (fig.~\ref{fig:data2}). The agreement of
$\omega_h$ and $\varphi$ with the theoretical predictions is
remarkable in view of the very weak velocity signal, providing
strong evidence that the weak secondary flow that we observe
originates from the precession of the experiment by the Earth
rotation. The magnitude of the secondary rotation lies in the
range $(1.5-3) \times 10^{-3}\,\Omega_0$, confirming that the
rotation vector ${\boldsymbol \omega}$ of the fluid is almost
aligned with ${\bf \Omega}_0$, with a very weak angular departure
of $\omega_h / \Omega_0 < 0.2$\degre.

\section{Conclusion}

Measuring the influence of the Earth rotation at the laboratory
scale is a technical challenge. In the fluid analogue of the
Foucault pendulum presented in this letter, the very weak
precession driven flow would have been impossible to detect
directly from the laboratory frame. Probing the flow in the
rotating frame naturally subtracts the first-order rotation and
allows us to detect this slight correction. We note that such
residual tilt-over flow forced by the Earth rotation defines an
irreducible background flow which should be present in every
rotating fluid experiments, routinely used as models for
geophysical and astrophysical flows in the laboratory.

\acknowledgments

We acknowledge C. Lamriben and M. Rabaud for fruitfull
discussions, and A. Aubertin, L. Auffray, C. Borget, A. Campagne
and R. Pidoux for experimental help. J.~B. is supported by the
``Triangle de la Physique''. This work is supported by the ANR
through grant no. ANR-2011-BS04-006-01 ``ONLITUR''. The rotating
platform ``Gyroflow'' was funded by the ``Triangle de la
Physique''.

\end{document}